\documentclass[aip,prx,twocolumn,showpacs,superscriptaddress,groupedaddress]{revtex4}  
\usepackage{graphicx}  
\usepackage{dcolumn}   
\usepackage{bm}        
\usepackage{amssymb}   
\usepackage{amsmath}
\usepackage{xfrac}

\begin{document}


\title{Magnetotransport Experiments on Fully Metallic Superconducting Dayem Bridge Field-Effect Transistors}
\author {Federico Paolucci}
\email{federico.paolucci@pi.infn.it}
\affiliation{INFN Sezione di Pisa, Largo Bruno Pontecorvo, 3, I-56127 Pisa, Italy}
\affiliation{ NEST, Instituto Nanoscienze-CNR and Scuola Normale Superiore, I-56127 Pisa, Italy}
\author {Giorgio De Simoni}
\affiliation{ NEST, Instituto Nanoscienze-CNR and Scuola Normale Superiore, I-56127 Pisa, Italy}
\author {Paolo Solinas}
\affiliation{ SPIN-CNR, Via Dodecaneso 33, I-16146 Genova, Italy}
\author {Elia Strambini}
\affiliation{ NEST, Instituto Nanoscienze-CNR and Scuola Normale Superiore, I-56127 Pisa, Italy}
\author {Nadia Ligato}
\affiliation{ NEST, Instituto Nanoscienze-CNR and Scuola Normale Superiore, I-56127 Pisa, Italy}
\author {Pauli Virtanen}
\affiliation{ NEST, Instituto Nanoscienze-CNR and Scuola Normale Superiore, I-56127 Pisa, Italy}
\author {Alessandro Braggio}
\affiliation{ NEST, Instituto Nanoscienze-CNR and Scuola Normale Superiore, I-56127 Pisa, Italy}
\author {Francesco Giazotto}
\affiliation{ NEST, Instituto Nanoscienze-CNR and Scuola Normale Superiore, I-56127 Pisa, Italy}

\begin{abstract}
In the last 60 years conventional solid and electrolyte gating allowed sizable modulations of the surface carrier concentration in metallic superconductors resulting in tuning their conductivity and changing their critical temperature. Recent conventional gating experiments on superconducting metal nano-structures showed full suppression of the critical current without variations of the normal state resistance and the critical temperature. These results still miss a microscopic explanation. In this article, we show a complete set of gating experiments on Ti-based superconducting Dayem bridges and a suggested classical thermodynamic model which seems to account for several of our experimental findings. In particular, zero-bias resistance and critical current $I_C$ measurements highlight the following: the suppression of $I_C$ with both polarities of gate voltage, the surface nature of the effect, the critical temperature independence from the electric field and the gate-induced growth of a sub-gap dissipative component. In addition, the temperature dependence of the Josephson critical current seems to show the transition from the ballistic Kulik-Omelyanchuck behavior to the Ambegaokar-Baratoff tunnel-like characteristic by increasing the electric field. Furthermore, the $I_C$ suppression persists in the presence of sizeable perpendicular-to-plane magnetic fields. We propose a classical thermodynamic model able to describe some of the experimental observations of the present and previous works. Above all, the model grabs the bipolar electric field induced suppression of $I_C$ and the emergence of a sub-gap dissipative component near full suppression of the supercurrent. Finally, applications employing the discussed effect are proposed.
\end{abstract}

\pacs{}
\maketitle

\section{Introduction}

Metals are believed to be insensitive to field-effect, because an external electric field vanishes within a depth comparable to the Thomas-Fermi length (smaller than the atom dimension). However, there is no physical reason to exclude the possibility to modulate the conductivity of metallic thin films via charge accumulation or depletion. Starting from the 50s several gating experiments showed the impact of electrostatic charging on the conductivity of metallic thin films \cite{Bonfiglioli1956, Bonfiglioli1959, Berman1975} and on the transition temperature of metallic superconductors \cite{Glover1960, Bonfiglioli1962}. Furthermore, recent calculations about gating on a metallic superconductor (Pb) showed that the electric field penetrates for a maximum depth of a few times the Thomas-Fermi length and the perturbation to superconductivity extends into the bulk of the superconductor for at least one coherence length \cite{Ummarino2017}. Conventional solid gating allows to realize electric fields of the order of maximum $10^9$ V/m. As a consequence, the maximum variation of free carrier concentration is about a few percent within the penetration depth of the electric field that is reflected in a modulation of conductivity of the same order of magnitude \cite{Berman1975}. Analogously, superconducting metallic thin films showed a change of the critical temperature $T_C$ of a few percent when exposed to strong electric fields \cite{Glover1960}. In particular, charge depletion (negative field) caused an increase of normal-state resistance and decrease of critical temperature, while carrier accumulation (positive field) originated lower resistance and higher critical temperature. The advent of electrolyte gating \cite{Panzer2005}, which relies on the voltage-induced polarization of an electrolyte (which is an ionic conductor and an electronic insulator) between a counter electrode and the sample, allowed to achieve electric fields as large as $10^{10}$ V/m and surface carrier modulations of $10^{15}$ charges cm$^{-2}$ for gate voltages of the order of a few volts \cite{Dhoot2006}. Therefore, conductivity modulations of about $10\%$ have been demonstrated in metals \cite{Daghero2012, Tortello2013}, such as gold, copper and silver, and sizable changes of the critical temperature of $\sim1\%$ in metallic superconductors, such as for niobium \cite{Choi2014} and niobium nitride \cite{Piatti2017}, have been achieved. Recent experiments investigated the dependence both of critical current $I_C$ and critical temperature of metallic superconductors (aluminum and titanium) on conventional solid gating \cite{DeSimoni2018, Paolucci2018, Varnava2018}. These studies focused on electric fields reaching maximum intensities of the order of $10^8$V/m, where the variation of charge carrier concentration in metals is negligibly small. As a matter of fact, on the one hand the normal-state resistance and the superconducting transition temperature were independent of the applied gate voltage, on the other hand the critical current was suppressed for both polarities of the electric field \cite{DeSimoni2018, Paolucci2018}. These experimental results seem to exclude both charge accumulation/depletion and quasiparticles overheating at the origin of the supercurrent suppression, but a microscopic explanation of the phenomenon is still missing.

\begin{figure}
\includegraphics {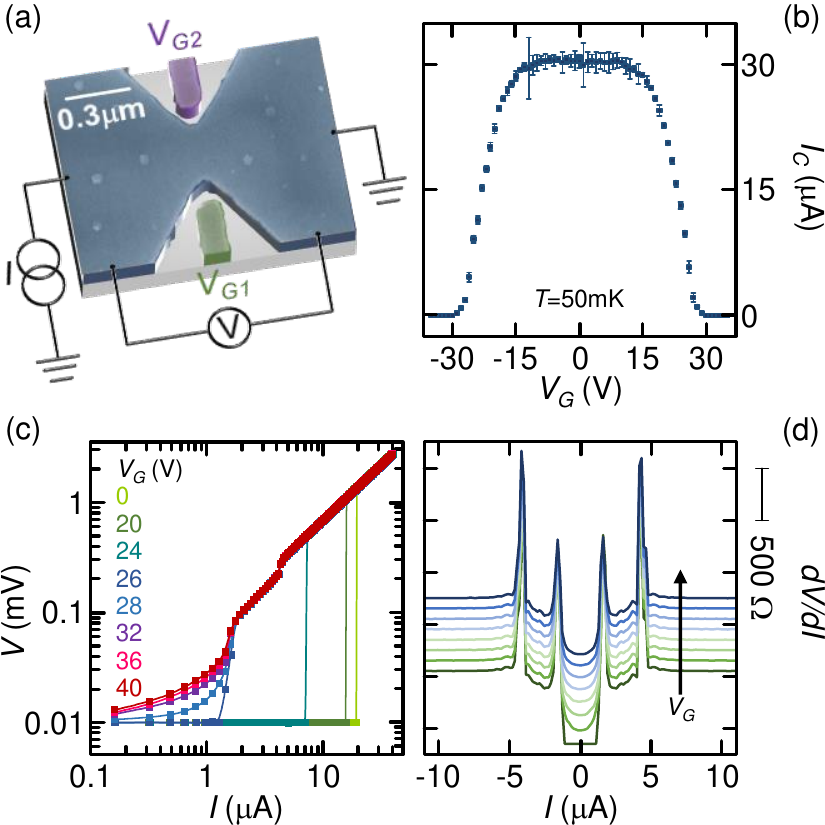}
\caption{\label{fig:Fig1} (a) Schematic representation of a typical $FET$ device. The Josephson junction (blue) is current biased and the voltage drop is measured with a room temperature voltage preamplifer, while the gate voltages $V_{G1}$ and $V_{G2}$ are applied to down (green) and up (violet) lateral gate electrode, respectively. (b) Dependence of the critical current $I_C$ on the gate voltage applied to both electrodes $V_G=V_{G1}=V_{G2}$ at a bath temperature $T=50$mK. (c) Log-log current-voltage ($I-V$) characteristics of a $DB-FET$ measured at $T=50$mK for different values of gate voltage $V_G=V_{G1}=V_{G2}$. (d) Dependence of $dV/dI$ on injection current $I$ for $V_G=V_{G1}=V_{G2}=26-40$V with step size of 2V. The experimental traces are vertically shifted for clarity.}
\end{figure}

Here, we show an extensive set of solid gating experiments performed on titanium-based Dayem bridge field-effect transistors $DB-FET$s and a superconducting thermodynamic model which seems able to support the phenomenology typical of several experimental observations presented in this work and in the previous studies \cite{DeSimoni2018, Paolucci2018}. The article is organized as follows: Section \ref{Basic} presents a basic characterization of the electric field dependent suppression of $I_C$ and the study of the joined impact of two different electric fields on the critical current; Section \ref{Joseph} shows the evolution of Josephson effect with gate voltage and its interpretation through a theory accounting for the change of the transmission probability of the junction; Section \ref{Temperature} displays the temperature dependence of the resistance measured for different values of injection current and gate voltage; Section \ref{Magnetic} describes the joined impact of electric and magnetic fields on both critical temperature and critical current of the Dayem bridge; Section \ref{Theo} presents a classical thermodynamical that seems able to explain some of the observations such as the gate voltage induced suppression of the critical current and the invariance of the superconducting transition temperature; and Section \ref{Possible applications} describes in detail two possible applications of our devices in the framework of both quantum and classic computation. 

\section{Basic Characterization of Field-Effect}\label{Basic}
Fig.\ref{fig:Fig1}-(a) shows a schematic representation of a $DB-FET$, where the top surface is a scanning electron micrograph of a real device and the electrical connections are represented. A typical gate-tunable Dayem bridge Josephson junction consists of a titanium (Ti) strip ($\sim4\mu$m wide and $\sim30$nm thick) interrupted by a constriction ($\sim125$nm long and $\sim300$nm wide). Two side electrodes [green and violet stripes in Fig. \ref{fig:Fig1}-(a)] placed at a distance of $\sim 80-120$nm from the constriction allow to independently apply two different gate voltages $V_{G1}$ and $V_{G2}$ and, therefore, electric fields on the Josephson junction ($JJ$) region. The $DB-FET$s were nano-fabricated through single step electron beam lithography ($EBL$) followed by evaporation of Ti on top of a p$^{++}-$doped silicon substrate covered with silicon dioxide (300nm thick). The titanium thin films were deposited in the ultra-high vacuum chamber (of base pressure on the order of $10^{-11}$Torr) of an electron beam evaporator at a rate $\sim13$\AA/s. The magneto-electric characterization of the $DB-FET$s was performed in a filtered $^3$He-$^4$He dry dilution refrigerator (two stage $RC-$ and $\pi-$filters) using standard four-wire technique. A DC current was injected using a low-noise source, the voltage across the bridge was measured by room-temperature preamplifiers, and gate voltage was applied using a source measure unit through different room temperature low-frequency filters ($\tau\sim1-100s$) with identical results.

Measurements of the superconducting critical current $I_C$ as a function of the same voltage applied simultaneously to both gate electrodes ($i.e.$ $V_G=V_{G1}=V_{G2}$) at a bath temperature $T=50$mK are shown in Fig. \ref{fig:Fig1}-(b). In full agreement with the first demonstrations of electric field dependent modulation of critical current in metallic $BCS$ superconductors \cite{DeSimoni2018, Paolucci2018}, $I_C$ is almost unaffected for low values of $V_G$ and starts to monotonically reduce till its full suppression at a critical voltage $V_{GC}\simeq26$V ($I_C$ is suppressed for both positive and negative values of the gate voltage). Notably, the field-effect induced reduction of critical current in our $DB-FET$s is \textit{symmetric} in the polarity of $V_G$. Since the intrinsic value of the chemical potential of the different stable phases of titanium does not reside at the van Hove singularity in the electron density of states \cite{Jafari2012, Zavodinsky2018} and the maximum superconducting correlations occurs at the van Hove singularity point, we can conclude that charge filling/depletion \cite{Choi2014, Piatti2017, Clark1980, Takayanagi1985, Akazaki1996} cannot be at the origin of such a behavior (the critical current would increase/decrease for charge filling/depletion or vice versa). 

Fig. \ref{fig:Fig1} (c) displays the $I-V$ characteristics of a Dayem bridge field-effect transistor measured for different values of gate voltage ($V_G=V_{G1}=V_{G2}$) at a temperature $T=50$mK plotted in double-logarithmic scale. On the one hand, for $V_G\leq26$V, the current flow is dissipationless ($V=0$) under the critical current, while above $I_C$ the $I-V$ characteristics is completely ohmic. On the other hand, for $V_G\geq28$V, charge transport is always dissipative ($V>0$), but the ohmic behavior ($V\propto I$) starts at $I\simeq5\mu$A independently of the value of gate voltage; furthermore, the emergence of a small jump in the $I-V$ characteristics at about 4$\mu$A is probably related to the transition into the normal state of a small portion of the device less affected by the electric field. The plots in Fig. \ref{fig:Fig1} (c) resemble the behavior of the critical current in an overdamped Josephson junction near the critical temperature in the presence of a sizeable thermal noise \cite{Ambegaokar1969}. Since our measurements are performed at about $0.1T_C$ ($T_C\simeq 540$mK) and there is no evidence of gate-dependent thermal (or even electrical) noise, we excluded such a mechanism as the origin of the presented effect.

\begin{figure}
\includegraphics {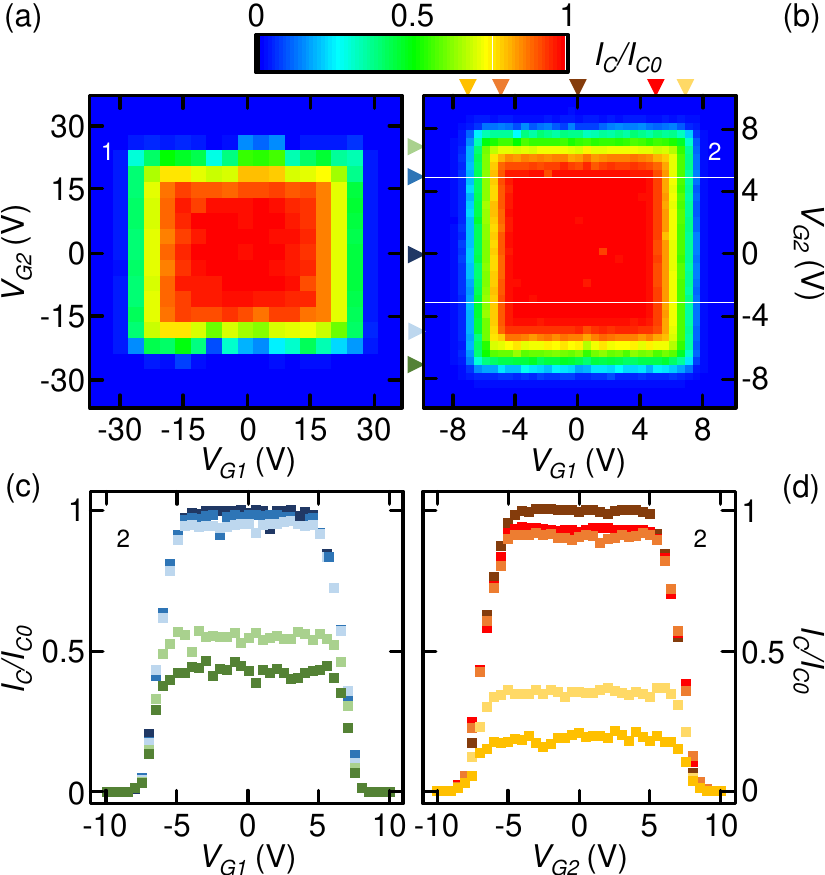}
\caption{\label{fig:Fig2} (a),(b) Color plot of the normalized critical current $I_C/I_{C0}$ as a function of $V_{G1}$ and $V_{G2}$ measured at a bath temperature $T=50$mK for two different devices ($1$ and $2$). (c) $I_C/I_{C0}$ vs $V_{G1}$ for different values of $V_{G2}$ indicated by the arrows in plot (b). (d) $I_C/I_{C0}$ vs $V_{G2}$ for different values of $V_{G1}$ indicated by the arrows in plot (b).}
\end{figure}

The dissipative behavior in the transistors is emphasized by plotting the differential resistance $dV/dI$ as a function of injection current $I$ for different values of the gate voltage around its critical value $V_{GC}=28$V [see Fig. \ref{fig:Fig1}-(d)]. In particular, $dV/dI$ shows a plateau under $I_C$ only for $V_G=26$V, while for higher values of gate voltage the dissipative contribution at low injection current rises. Since the data are the numerical derivative of measurements acquired by sweeping the injection current from negative to positive values, the peaks for $I<0$ represent the retrapping current $I_R$, whereas the spikes for $I>0$ indicate the critical current $I_C$. Notably, $I_R$ stays almost constant in the range of applied gate voltages. The traces in Fig. \ref{fig:Fig1}-(d) show two main features. First, the Dayem bridge exhibits two transitions for values of gate voltage approaching $V_{GC}$ [see also Fig. \ref{fig:Fig1}-(c)]. Second, only for $V_G>26$V the critical current and the retrapping current of both transitions are equal ($I_R=I_C$, because the critical current lowers and the retrapping current cannot exceed the value of $I_C$), as reported in previous gating experiments on $BCS$ wires \cite{DeSimoni2018}. 

\begin{figure*}
\includegraphics {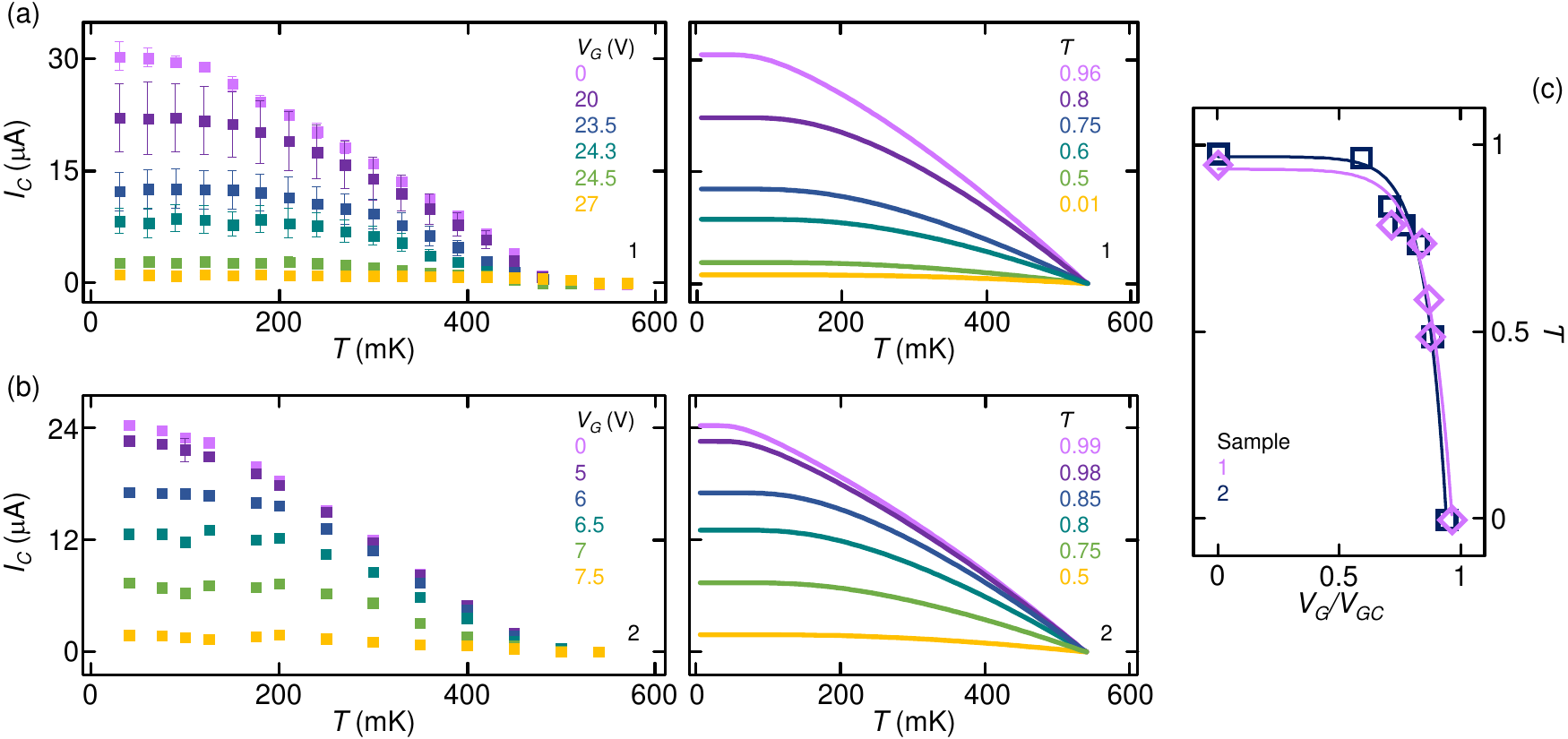}
\caption{\label{fig:Fig5} (a),(b)  Left panel: Critical current $I_C$ as a function of the bath temperature $T$ for different values of gate voltage $V_G=V_{G1}=V_{G2}$ measured on sample $1$ and $2$, respectively. Right panel: Fit of the behavior of I$_C$ vs $T$ calculated with the Kulik-Omelyanchuk theory generalized for an arbitrary effective transmissivity $\mathcal{T}$ of the Josephson junction \cite{Golubov2004}. The extracted values of $\mathcal{T}$ correspond to the gate voltages represented with the same color in left panel. (c) Effective transmissivity $\mathcal{T}$ of the $JJ$ vs normalized gate voltage $V_G/V_{GC}$ extracted for the devices  1 (purple) and 2 (blue).}
\end{figure*}

To study the electric field impact on a superconducting metal and its effect on $I_C$, we measured the critical current while applying two independent voltages $V_{G1}$ and $V_{G2}$ to the side gate electrodes. The contour plots in Fig. \ref{fig:Fig2}-(a),(b) show the normalized critical current $I_C/I_{C0}$ (where $I_{C0}$ is the critical current with no gate voltage applied) as a function of both $V_{G1}$ and $V_{G2}$ measured for two different $DB-FET$s ($1$ and $2$) at $T=50$mK. The gate electrodes in sample $1$ are placed at a distance of about $120$nm from the constriction, while in sample $2$ they are separated by a gap of $\sim80$nm. The color plots of our $DB-FET$s show a square-like shape [the rectangular shape in  Fig. \ref{fig:Fig2}-(a) stems from the different distance of the two gates from the active region] that indicates the \textit{independence} of the effects on the critical current of $V_{G1}$ and $V_{G2}$. This behavior is highlighted by the plots of $I_C/I_{C0}$ as a function of $V_{G1}$ ($V_{G2}$) for fixed values of $V_{G2}$ ($V_{G1}$) shown in Fig. \ref{fig:Fig2}-(c),(d) for the contour plot of Fig. \ref{fig:Fig2}-(b). On the one hand, the traces confirm the independence of the critical voltage of one gate from the other. On the other hand, the plots highlight a similar reduction effect of $V_{G1}$ and $V_{G2}$ on $I_C$. The small asymmetry existing with the sign of $V_{G1}$ and $V_{G2}$ arises from the experimental procedure followed to carry out the experiments: both gate voltages were swept from negative to positive values. As a consequence, due to a small hysteresis in the electric field response, the critical current is slightly smaller for negative values of the gate voltage [$I_C(-V_G)\lesssim I_C(V_G)$].

The phenomenology related to the dependence of the effect of $V_{G1}$ and $V_{G2}$ on $I_C$ seems to suggest that the suppression of critical current is related to a surface effect, which nonlocally affects the superconductivity within a distance of a few times of the superconducting coherence length $\xi$, as recently calculated for another $BCS$ superconductor (Pb) \cite{Ummarino2017}. Specifically, in similar experimental conditions of the present measurements, i.e., in solid-gated Ti wires far below $T_C$, the electric field induced suppression of critical current has been shown to extend into the superconductor bulk for a few times the coherence length \cite{DeSimoni2018}.

\section{Josephson effect}\label{Joseph}

The Dayem bridge geometry chosen for our devices allows to study the dependence of the Josephson coupling on the applied gate voltage $V_G=V_{G1}=V_{G2}$. The left panels of Fig. \ref{fig:Fig5}-(a),(b) show the dependence of the critical current $I_C$ on the bath temperature $T$ measured for different values of $V_G$. By increasing the gate voltage, on the one hand the critical current is suppressed at every temperature in the measured range ($T<600$mK), while on the other hand the Josephson supercurrent vanishes always at the same bath temperature $T$, i.e., the temperature when no gate bias is applied ($T_C\simeq540$mK). We observe that the suppression of $I_C$ in a Josephson junction due to quasiparticles injection generated by $V_G$ would increase the effective electronic temperature $T_{eff}$ \cite{Morpurgo1998}. Therefore, the Josephson supercurrent would vanish ($I_C=0$) when $T_{eff}=T_C$, but $T<T_C$. As a consequence, we can reasonably exclude quasiparticle overheating as the source of critical current suppression in our $DB-FET$s. 

A further analysis of the data shows that $I_C$ of the pristine $JJ$ (i.e., with $V_G=0$) has the typical Kulik-Omelyanchuck \cite{Kulik1977} dependence on temperature characteristic of a clean ballistic constriction, while near full Josephson supercurrent suppression ($V_G \simeq V_{GC}$) the device has the Ambegaokar-Baratoff \cite{Ambegaokar1963} behavior as  a tunnel $JJ$  [see the left panels of Fig. \ref{fig:Fig5}-(a),(b)].

The critical current of a Josephson constriction in the short junction limit can be written \cite{Golubov2004}:
\begin{equation}
\begin{split}
I_C(\varphi)=\frac{\pi\Delta}{2eR_N}\frac{\sin\varphi}{\sqrt{1-\mathcal{T}\sin^2\frac{\varphi}{2}}}\\
\times\tanh\left[ \frac{\Delta}{2T}\sqrt{1-\mathcal{T}\sin^2\frac{\varphi}{2}}\right],
\end{split}
\label{eq:CritCurr}
\end{equation}
where $\varphi$ is the phase difference across the $JJ$, $\Delta$ is the pairing potential, $e$ is the electron charge, $\mathcal{T}$ is the transmission probability and $R_N$ is the contact normal-state resistance. The latter can be written as $R_N=\sfrac{R_{Sh}}{\mathcal{T}}=\sfrac{(4\pi^2\hbar)}{(e^2k_F^2S\mathcal{T}})$, with $R_{Sh}$ the Sharvin resistance, $k_F$ the Fermi wave vector and $S$ the constriction cross-sectional area. On the one hand, for $\mathcal{T}=1$, Eq. \ref{eq:CritCurr} adheres to the Kulik-Omelyanchuck advanced theory \cite{Kulik1977}. On the other hand, for $\mathcal{T}\ll1$ it reduces to the Ambegaokar-Baratoff model \cite{Ambegaokar1963}. 

We fit the experimental data using Eq.\ref{eq:CritCurr}. The right panels of Fig. \ref{fig:Fig5}-(a),(b) show the theoretical traces calculated for the experimental curves in left panels (the values of $\mathcal{T}$ correspond to $V_G$s represented with the same color). We would like to stress that in our experiments the normal-state resistance is unaffected by the electric field. To use this model regardless, we take a phenomenological point of view and assume that some channels do not contribute to the supercurrent. The microscopic mechanism how this would occur is however unclear at present. However, the fits show a good agreement with the experimental data. In particular, the experimental curves measured for increasing gate voltage correspond to theoretical curves calculated for decreasing effective transmission probability. The theory is not able to grasp is the change of concavity in the $I_C$ vs $T$ trace near the critical temperature when a gate voltage is applied. This model considers a temperature-independent transmission probability. However, since the effectiveness of the critical current suppression in similar structures has been shown to decrease by increasing the temperature \cite{DeSimoni2018, Paolucci2018}, we can speculate that the value of $\mathcal{T}$ arisng from the analysis of the present data should be temperature-dependent (especially near $T_C$).

To compare $FET$ devices with gate electrodes placed at different distances from the $JJ$ we plot $\mathcal{T}$ as a function of the normalized gate voltage $V_G/V_{GC}$ [see Fig. \ref{fig:Fig5}-(c)]. Despite the critical voltages of the two transistors are extremely different ($V_{GC}^1\simeq4V_{GC}^2$ with $V_{GC}^1$ and $V_{GC}^2$ critical voltage of device 1 and 2, respectively), the effective transmission probability seems to show an universal dependence on the normalized gate voltage. In particular, $\mathcal{T}$ drastically reduces in proximity of $V_{GC}$ suggesting the gate-induced creation of dissipative regions into the Dayem Bridge not able to carry the Josephson supercurrent. Furthermore, the dissipative states observed in the $I-V$ characteristics at high values of $V_{G}$ [as shown in Fig. \ref{fig:Fig1}-(c)] seems to indicate the formation of disordered puddle-like normal metal-superconducting regions with a non-continuous superconducting path.

Summarizing, the entire phenomenology of the Josephson critical current seems to be compatible with the growth of dissipative regions in the Dayem bridge constriction by increasing the gate voltage and the creation of a percolative path for the Josephson current translated in the reduction of the effective transmission probability of the junction.

\section{Temperature Dependence of Resistance} \label{Temperature}

\begin{figure}
\includegraphics {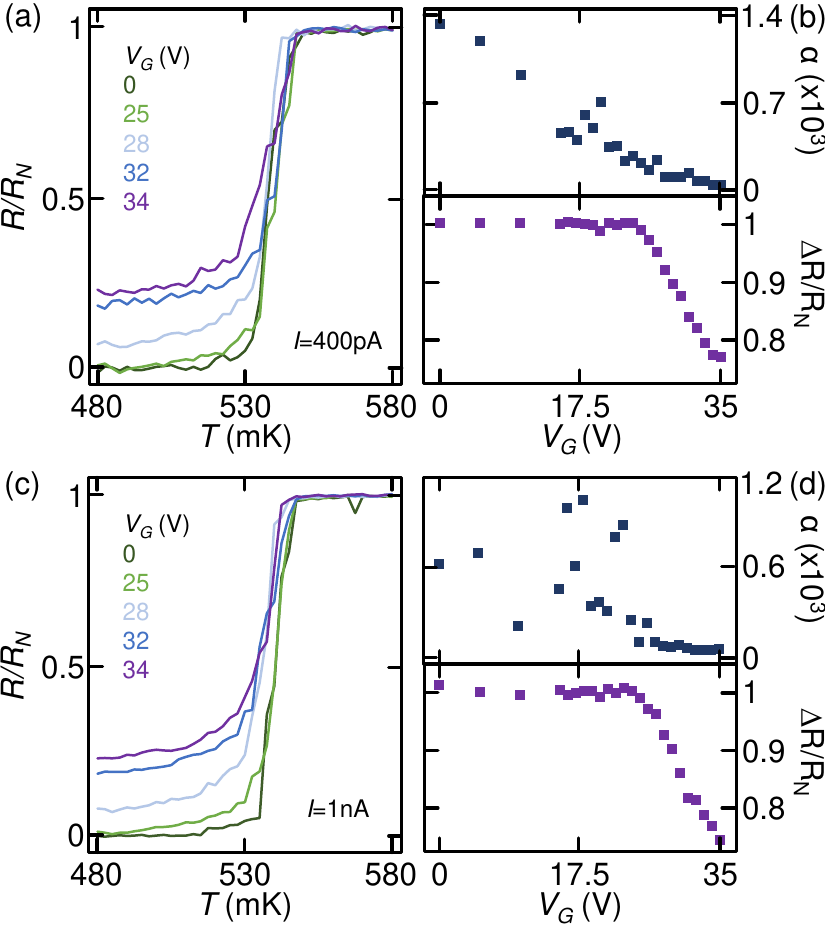}
\caption{\label{fig:Fig3} (a),(c) Normalized zero-bias resistance $R/R_N$ as a function of temperature $T$ measured in sample $1$ for different values of the gate voltage $V_G=V_{G1}=V_{G2}$ with 400pA (a) and 1nA (c) current bias. (b),(d) Electrothermal parameter $\alpha$ (top panel) and normalized resitance jump at the superconductor/normal metal transition $\Delta R/ R_N$ (bottom panel) as a function of gate voltage $V_G=V_{G1}=V_{G2}$ for 400pA (a) and 1nA (c) current bias.}
\end{figure}

The resistance vs temperature characteristics of the $FET$s were obtained by low frequency lock-in technique for different values of the gate voltage applied by a low-noise source-measure-unit. The voltage drop was amplified through room-temperature differential amplifiers. To evaluate the interplay of the bias current with the gate voltage on the superconductor-to-normal metal transition we repeated the experiments for two values of the current, $I=400$pA and $I=1$nA. In both cases we measured a resistance in the normal state $R_N\simeq600\Omega$, in agreement with the $I-V$ characteristics in Fig. \ref{fig:Fig1}-(c). Figure \ref{fig:Fig3}-(a),(c) show the temperature dependence of the normalized resistance $R/R_N$ for selected values of the gate voltage measured with $I=400$pA and $I=1$nA, respectively. In agreement with previous reports on gating of metallic wires \cite{DeSimoni2018} and Dayem bridges \cite{Paolucci2018}, the critical temperature stays constant within the experimental error for all the studied values of $V_G$. On the other hand, a sub-gap dissipative component starts to grow at high values of gate voltage for both injected currents [see Fig. \ref{fig:Fig3}-(a),(c)]. 

In order to highlight the dependence of the resistance on both temperature and electric field and to give a quantitative analysis of the data we introduce two dimensionless figures of merit: the \textit{electrothermal} parameter $\alpha$ \cite{Harwin2017} and the normalized resistance jump between superconducting and normal state.

The parameter $\alpha=\frac{T}{R(T, I)}\text{ }  \frac{\mathrm d R(T, I)}{\mathrm d T}$, where the resistance $R$ depends on both temperature $T$ and bias current $I$, characterizes the sharpness of the superconducting transition. The top panels of Fig. \ref{fig:Fig3}-(b),(d) show the dependence of the electrothermal parameter $\alpha$ calculated at half of the superconductor-normal metal transition (i.e. when $R=R_N/2\simeq 300\Omega$) as a function of gate voltage for $I=400$pA and $I=1$nA, respectively. For all values of $V_G$, the transition is sharper (i.e. the parameter $\alpha$ assumes higher values) for low applied bias, because depairing probability increases with biasing current \cite{Levine1965}. By increasing gate voltage $\alpha$ lowers, because the superconducting transition widens in temperature and decreases in height [see Fig. \ref{fig:Fig3}-(a),(c)]. The electrothermal parameter shows strong fluctuations with $V_G$ in correspondence of the beginning of the critical current suppression [see Fig. \ref{fig:Fig1}-(b)]. At these values of gate voltage the noise of the critical current rises \cite{DeSimoni2018, Paolucci2018}, because the electric field apparently yields more instabilities to the superconducting condensate. The latter is reflected in a randomization of the superconducting transition and a fluctuating parameter $\alpha$ especially for higher bias current (where the depairing is naturally stronger).  

\begin{figure*}
\includegraphics {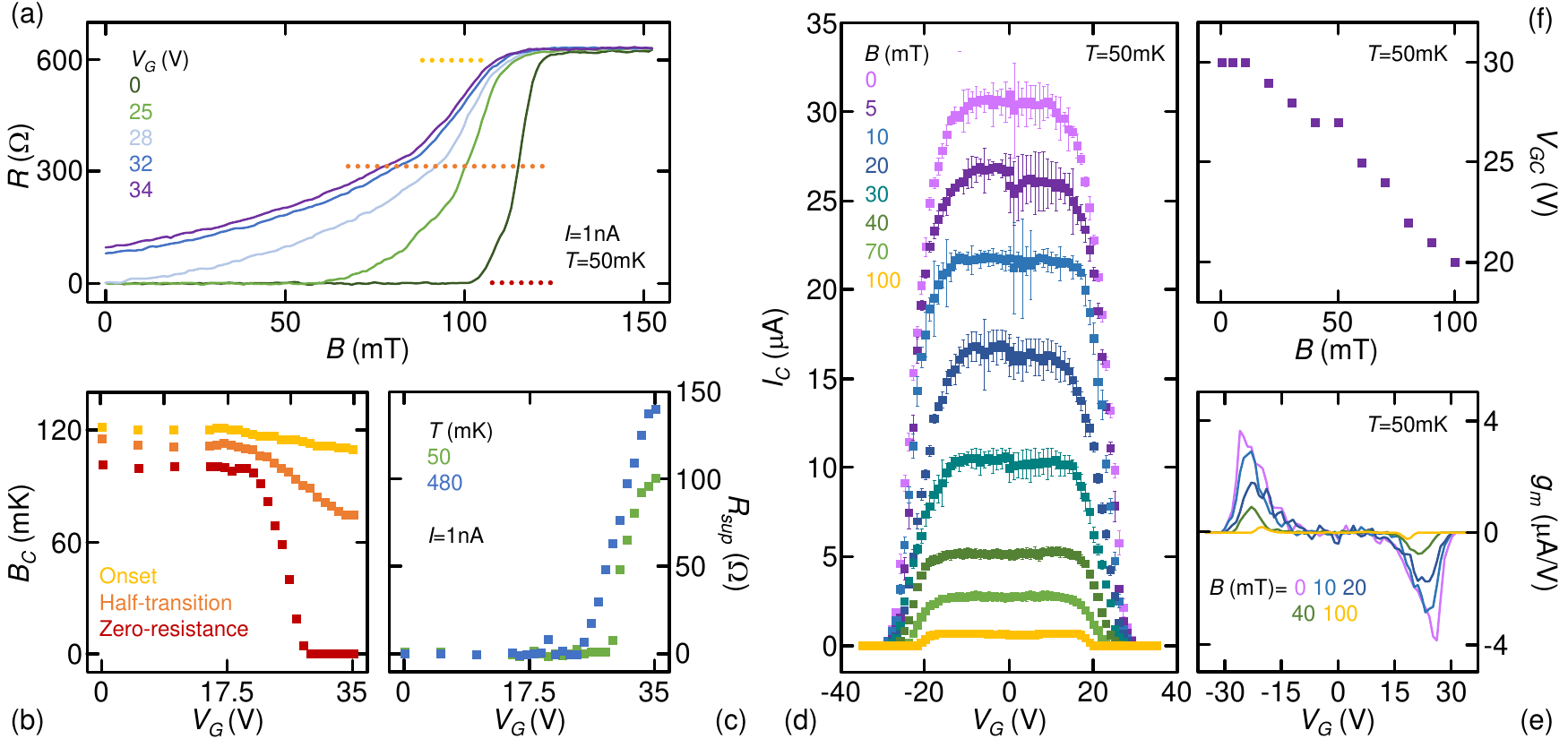}
\caption{\label{fig:Fig4} (a) Zero-bias resistance $R$ as a function of perpendicular magnetic field $B$ measured at $T=$50mK for different values of gate voltage $V_G=V_{G1}=V_{G2}$ with 1nA current bias. (b) Critical magnetic field $B_C$ extracted from $R$ vs $B$ mesurements (a) at the oneset  (yellow), half (orange) and complete (red) transition to the superconducting state. (c) Resistance in the superconducting state $R_{sup}$ as a function of gate voltage $V_G=V_{G1}=V_{G2}$ measured at $T=50$mK [green, extracted from data of panel (a)] and $T=480$mK [blue, extracted from data of Fig. \ref{fig:Fig3} (c)] with 1nA current bias. (d) $I_C$ vs $V_G=V_{G1}=V_{G2}$ measured at $T=50$mK for different values of perpenducular-to-plane magnetic field $B$. (e) Transconductance $g_m$ as a function of gate voltage $V_G=V_{G1}=V_{G2}$ measured at $T=50$mK for different values of $B$. (f) Critical gate voltage $V_{GC}$ as a function of perpendicular magnetic field $B$ measured at $T=50$mK.}
\end{figure*}

The normalized resistance jump across the superconducting transition is defined $\sfrac{\Delta R}{ R_N}=\sfrac{[R(T=480mK)-R_N]}{R_N}$ and is a way to quantify the zero-bias dissipative component growing with gate voltage [see lowe panels of Fig. \ref{fig:Fig3}-(b),(d)]. The latter starts to decrease at $V_G\simeq25$V for both the injection currents. By further increasing gate voltage the normalized resistance jump decreases almost linearly down to $\sim0.75$ for $V_G=35V$ [where $I_C=0$, see Fig. \ref{fig:Fig1}-(b)] with a small difference (a few percent) between the two bias currents. The growth of the zero-bias dissipative component below the critical temperature (in the following denoted with $R_{sup}$) once more seems to suggest the creation and manifestation of normal metal regions interrupting the superconductor and establishing a percolative pattern for the Josephson supercurrent. Furthermore, the invariance of $R_{sup}$ with the current seems to indicate that the remaining superconducting regions can support a higher supercurrent, as evident by looking at the low temperature $I-V$ characteristics shown in Fig. \ref{fig:Fig1}-(c).

\section{Joined impact of electric and magnetic field}\label{Magnetic}
The interplay between electric field and magnetic field was investigated both by zero-bias experiments ($R$ vs $B$ at fixed $V_G$) and by critical current measurements ($I_C$ vs $V_G$ at fixed $B$). The experimental setups are already presented in Sections \ref{Basic} and \ref{Temperature}, while the magnetic field is applied out-of-plane (perpendicular to the superconducting film). 

We first present the evolution of the critical magnetic field $B_C$ with gate voltage $V_G$ measured with an injection current $I=1$nA at a bath temperature $T=50$mK [see Fig. \ref{fig:Fig4}-(a)]. On the one hand, the onset of the superconducting transition seems to slightly change with increasing gate voltage. On the other hand, the width of the transition increases with gate voltage until a resistive component grows even at $B=0$ [in agreement with the $R$ vs $T$ experiments shown in Fig. \ref{fig:Fig3}-(a),(c)]. 

To quantify the evolution of the critical magnetic field with the gate voltage we extrapolated the values of $B_C$ at the onset, half and full transition, as shown in Fig. \ref{fig:Fig4}-(b). We define onset of the transition the values of magnetic field corresponding to a zero-bias resistance of $600\Omega$ [$B(R=600\Omega)$]. The onset of the transition stays almost constant until the gate voltage reaches the critical value $V_{GC}=18$V, then the critical field slowly and almost lineraly decreases of about the $10\%$ of its intrinsic value. The half-transition magnetic critical field, i.e. the value of $B$ for which $R=R_N/2\simeq 300\Omega$, monotonically lowers from $B\simeq115$mT at $V_G\leq18$V to $B\simeq74$mT at $V_G=35V$ (with a total variation of about $35\%$). Finally, the zero-resistance critical field drops very steeply with gate voltage (in the range 18-29V) from $B\simeq100$mT to 0. The total suppression of the zero-resistance critical field highlights the growth of a sub-gap dissipative component as already reported in zero-bias experiments performed at temperatures close to $T_C$ [see Fig. \ref{fig:Fig3}-(a),(c)]. 

The resistance in the superconducting state $R_{sup}$ increases with temperature for a fixed value of gate voltage, as shown in Fig. \ref{fig:Fig4}-(c). This effect can be interpreted in terms of strengthening of the depairing of the Cooper pairs and the increasing probability of transition to the normal state of small superconducting areas with rising temperature.

Summarizing, the measurements of zero-bias resistance as a function of magnetic field for different values of the gate voltage provide useful information about the origin of the Josephson critical current suppression. On the one hand, the (almost) invariance of the transition onset for large $V_G$ values suggests to exclude quasiparticles overheating as the origin of the suppression of superconductivity, because the transition onset is expected to lower more evidently by increasing the electronic temperature \cite{Zhang2013}. On the other hand,  the behavior of the three critical fields highlights the broadening of the normal metal-to-superconductor transition which increases from $\sim27$mT to  $\sim100$mT, that could be due to the formation of normal metal areas. Possibly these normal portions could affect the dependence of the zero-bias resistance on the magnetic field due to inverse proximity effect \cite{deGennes, tinkham2012introduction}, but the lack of a microscopic description of the effect impedes to quantify their real impact.

We now focus on the combined impact of electric and magnetic field on the critical current in the Dayem bridge field-effect transistors. Figure \ref{fig:Fig4}-(d) shows the dependence of $I_C$ on $V_G$ for different values of perpendicular-to-plane magnetic field. The critical current suppression is almost invariant with the polarity of electric field, the plateau of constant $I_C$ widens by increasing magnetic field and the critical voltage for full suppression of the Josephson supercurrent $V_{GC}$ lowers by rising $B$. These results resemble those reported on similar experiments performed on superconducting wires \cite{DeSimoni2018} that do not have a microscopic explanation yet (a phenomenological \textit{ad hoc} Ginzburg-Landau $GL$ model was exploited to make a comparison to the data \cite{DeSimoni2018}). The widening of the plateau with rising magnetic field suggests again to exclude quasiparticle overheating at the origin of supercurrent suppression, because the dependence of $I_C$ on $B$ is more steep for temperatures approaching $T_C$ \cite{Mydosh1965}. 

To highlight the decrease of the critical current modulation with the gate voltage for increasing values of the magnetic field we calculated the transconductance $g_m=dI_C/dV_G$ as a function of $V_G$ for different values of $B$, as plotted in Figure \ref{fig:Fig4}-(e). The transconductance lowers by more than one order of magnitude by increasing the magnetic field and its maximum value moves towards lower electric fields (both in positive and negative polarity) by increasing $B$. In particular, the maximum of transconductance varies from $g_m\simeq3.6\mu$A/V at $V_G=26$V for $B=0$ to $g_m\simeq0.2\mu$A/V at $V_G=18$V for $B=100$mT.  

The critical voltage is almost constant for small values of magnetic field and then decreases almost linearly with $B$ [see Fig. \ref{fig:Fig4}-(f)] from $V_{GC}\simeq30$V at $B=0$ to $V_{GC}\simeq20$V at $B=100$mT. The variation of $V_{GC}$ with the magnetic field is much stronger than with the temperature \cite{Paolucci2018} and resembles the behavior reported for solid gated superconducting metal wires \cite{DeSimoni2018}. All this phenomenology has no microscopic explanation yet.

\section{Thermodynamic model}\label{Theo}

The above experimental evidences and the recent works on electric field dependent supercurrent suppression in metallic superconductors \cite{DeSimoni2018, Paolucci2018} still miss a complete microscopic interpretation. Here, by starting with a simple working hypothesis we develop a classical thermodynamic model \cite{deGennes,tinkham2012introduction} accounting for several experimental observations. 

\begin{figure}
   \begin{center}
    \includegraphics{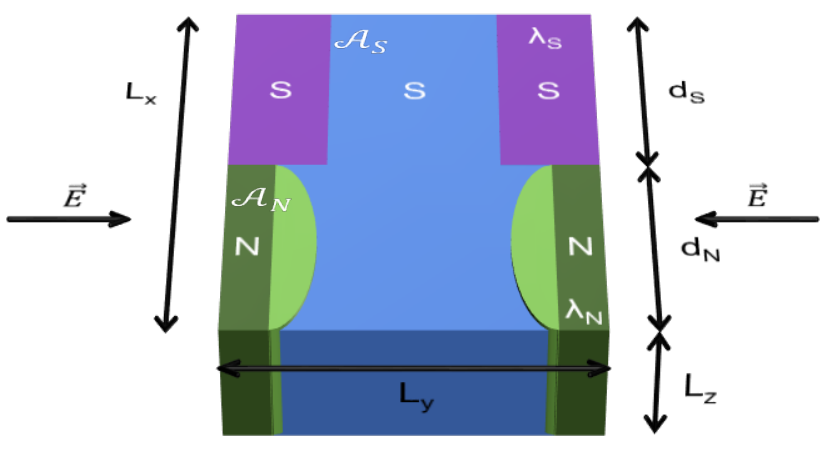}
    \end{center}
    \caption{Superconducting wire of dimensions $L_x$, $L_y$ and $L_z$ subject to a lateral electric field $\vec{E}$ along the $y$ direction. 
    $\lambda_S$ and $\lambda_N$ are the electric field penetration length into the superconducting $S$ (blue and violet) and normal region $N$ (dark and light green), respectively. $\mathcal{A}_S$ and $\mathcal{A}_N$ are the areas of the $S$ and $N$ regions, respectively.
    The boundary regions exposed to the electric field have lengths $d_S$ and $d_N$, respectively.
      }  
    \label{fig:wire}
\end{figure}

We consider a rectangular superconducting wire with dimensions $L_x$, $L_y$ and $L_z$ at temperature $T< T_C$ subject to a lateral electric field $\vec{E}$ pointing to the center of the sample, as shown in Fig. \ref{fig:wire}. Our theory is grounded on one key hypothesis: the electrostatic energy stored in a superconductor is assumed to be larger than that stored in a normal metal. This excess of electric energy in the superconductor could be due to an increase of the electric penetration length $\lambda_S$ and/or of the permittivity $\epsilon_S$, as theoretical studies on bulk superconductors \cite{Prange1963, Seiden1966, Machida2003} and experimental observations on thin films \cite{Jenks1993} seem to indicate. As a consequence, our working hypothesis can be reduced to assume that $\theta=\lambda_S \epsilon_S - \lambda_N \epsilon_N > 0$ where $\lambda_N$ and $\epsilon_N$ are the penetration length and the permittivity in the normal state, respectively. Notably, under this condition, superconductivity becomes \textit{unstable} when the excess electrostatic energy stored in the superconductor is comparable to its condensation energy. To simplify the discussion, we assume that the electric field is constant within the region penetrated by the electric field. The latter is a realistic hypothesis given the fact that the distance of the gate electrodes from the superconductor ($\sim 80-120$nm) is much larger than the electric field-penetrated region ($<1$nm \cite{Piatti2017, Ummarino2017}).

We denote with $F_S$ and $\mathcal{F}_S$ the free energy density and free energy associated to the superconductor, respectively.
We use analogous notations $F_N$ and $\mathcal{F}_N$ for the normal region.
By following the standard thermodynamic approach \cite{deGennes,tinkham2012introduction, Bao_NatComm2013} we can write the free energies associated to the superconducting and normal state as:
\begin{eqnarray}
  \mathcal{F}_{S} &=& V_S F_S + V_{S,E}^T u_{E,S} \nonumber \\
  \mathcal{F}_{N} &=& V_N F_N + V_{N,E}^T u_{E ,N}
  \label{eq:F_SN_energy}
\end{eqnarray} 
where $V_S$ is the superconductor volume, $V_N$ is the normal metal volume, while $V_{S,E}^T$ and  $V_{N,E}^T$ are the volumes of the superconductor and the normal metal penetrated by the electric field, respectively. Above, $u_{E,S} = \frac{1}{2} \epsilon_S E^2$ and $u_{E,N} = \frac{1}{2} \epsilon_N E^2$ are the  electric field energy densities. It is important to point out that from previous experiments \cite{Piatti2017, DeSimoni2018, Paolucci2018} and theoretical calculations \cite{Ummarino2017} it turns out that, in order to observe the electric field effect, the length along which we apply the electric field is at most a few times the coherence length. The present classical model does not account for this microscopic information. In the following we assume to be in these conditions. 

\subsection{Superconductor to normal metal transition}
Here, we focus on the case when all the wire is in the superconducting or in the normal state, thus, with respect to Fig. \ref{fig:wire} we assume that $d_N=0$ or $d_S=0$, i.e. $V_S=V$ or $V_N=V$ where $V=L_z L_x L_z$ is the total volume. Accordingly, the volumes of the superconductor or the normal metal penetrated by the electric field $\vec{E}$ can be written as $V_{S,E}^T = L_z L_x  (2 \lambda_S)$ or $V_{N,E}^T= L_z L_x  (2 \lambda_N)$, where the factor 2 stems for the electric field applied to both sides of the specimen.

In the presence of a static electric field, the superconducting state is energetically favorable when $\mathcal{F}_{N} > \mathcal{F}_{S}$.
Keeping the generic volume expressions, this inequality reads: 
\begin{equation}
 	\mathcal{U}_{E,S} - \mathcal{U}_{E,N} = \Delta \mathcal{U}_E < V \Delta F
	\label{eq:u_e_cond1}
\end{equation}
where $\mathcal{U}_{E,S}$ and $\mathcal{U}_{E,N}$ are the electric field energies in the superconductor and normal metal, respectively, and $\Delta F = F_N-F_S$ is the condensation energy density.
This equation determines the electrical critical energy for which superconductivity is destroyed.

At $T=0$, the condensation energy density can be written as $\Delta F = N_0 \Delta_0^2/2$ \cite{tinkham2012introduction} where $\Delta_0$ is the zero-temperature zero-field superconducting energy gap and $N_0$ is the density of states in the metal at the Fermi level.
From Eq. (\ref{eq:u_e_cond1}), we can define the critical electric field for the superconductor-to-normal-metal transition as:
\begin{equation}
 E^2 < E_C^2 =  \frac{L_y}{2 \theta} N_0 \Delta_0^2.
 \label{eq:E_critical_SN}
\end{equation}

In order to give a rough estimate of $E_C$ and compare it with the experimental values we use the parameters extracted from our measurements. The width of the devce can be extracted from the typical dimension of the constriction $L_y= 300~$nm. The titanium energy gap is given by $\Delta_0 = 1.764~k_B~T_C=82\mu$eV with $T_C\simeq540$mK, and its density of states is $N_0 = 1.35 \times10^{47}~$J$^{-1}~$m$^{-3}$ \cite{DeSimoni2018}. Since the permittivities for both the superconductor and normal metal are unknown and their difference si small \cite{Prange1963, Seiden1966, Machida2003}, we assume both of them to be equal to the vacuum, i.e., $\epsilon_S = \epsilon_N = \epsilon_0 = 8.85 \times 10^{-12}~$F~m$^{-1}$. An electric field penetrates a normal metal for a depth comparable to the Thomas-Fermi screening $\lambda_{T-F}$, therefore we have $\lambda_N =\lambda_{T-F}=0.5\times10^{-10}~$m. For a superconductor we speculate a penetration $\lambda_S =\delta \lambda_N$, with the unknown scaling parameter $\delta>1$ since we assume longer penetration in $S$ than in $N$. In previous works on linear-response of bulk superconductors the values of $\delta$ approach 1 \cite{Prange1963, Seiden1966}. Notice that the value of the critical electric field changes weakly with $\lambda_S$ because of the square root dependence in Eq. (\ref{eq:E_critical_SN}). 

We now compare the theoretical values of $E_C$ with the values extrapolated from the experimental data. By employing $\delta =1.1$ we get $E_C = 2.8 \times 10^8$V/m, while by using $\delta =1.01$ the critical electric field is $8.9\times 10^8$V/m. In sample 1, the critical gate voltage $V_{GC}\simeq30$V with gate electrodes at a distance $d\simeq120$nm from the constriction yields a critical electric field $E_C=V_{GC}/d\simeq2.5\times 10^8$V/m. In sample 2, the critical voltage is $V_{GC}\simeq10$V ($d\simeq80$nm) and, as a consequence, the critical field is $E_C\simeq1.25\times 10^8$V/m. We can conclude that the theoretical model can be compatible with our experiments given a small deviation from the screening length of about $1-10\%$, which is sufficient for the observed superconducting to normal metal transition.

\subsection{Inhomogeneous state}

The experimental data showed the growth of a sub-gap dissipative state induced by the application of the external electric field [see Fig.\ref{fig:Fig1}-(c), Fig.\ref{fig:Fig3}-(a),(c) and Fig.\ref{fig:Fig4}-(a),(c)] in agreement with previous results of gating experiments on metallic superconductors \cite{Paolucci2018}. In the following we show how our thermodynamic model can also account for the appearance of an inhomogeneous state (where the superconductor and the normal state coexist in the wire) inducing a dissipative component in the transport characteristics.

In the superconductor-normal metal ($S-N$) inhomogeneous state the total volume of the wire can be divided in superconducting and normal metal portions $V=V_S+V_N$. Therefore, Eq. (\ref{eq:F_SN_energy}) has to take into account the superconductor and normal state reduced volumes $V_S = L_z \mathcal{A}_S $ and $V_N = L_z \mathcal{A}_N $ where $\mathcal{A}_S$ and $\mathcal{A}_N$ are the superconducting and normal metal surfaces, respectively (see Fig. \ref{fig:wire}), and they are related by $\mathcal{A} = L_x L_y = \mathcal{A}_S +\mathcal{A}_N$ (i.e. the total area needs to be conserved). In particular, we assume that the region directly exposed to the electric field is divided in superconducting and normal metal areas $L_x=d_S+d_N$ where $d_S$ is the superconducting portion and $d_N$ is the dissipative section. As a consequence, the free energies for the superconducting and normal state can be written as $\mathcal{F}_{S} = V_S F_S + V_{S,E}^T u_{E,S}$ and $\mathcal{F}_{N} = V_N F_N + V_{N,E}^T u_{E,N}$, and the total energy associated to this configuration is $\mathcal{F}_{S} +\mathcal{F}_{N}$.

We also need to consider the free energy contribution at the $S-N$ interface since passing from a superconducting to a normal region, the order parameter $\psi$ must change \cite{deGennes}. According to the Ginzburg-Landau theory the latter gives a contribution to the free energy proportional to $\nabla  \psi$. We denote the wall energy associate to the interface with $ \mathcal{F}_W = L_z l \gamma$ where $l$ is the total length of the interface and $\gamma$ is the surface energy per unit area \cite{deGennes}. Therefore, the most general expression for the \textit{inhomogeneous} state free energy reads:
\begin{equation}
 \mathcal{F}_I = V_S  F_S + V_{S,E}^T u_{E,N} +  V_N  F_N + V_{N,E}^T u_{E,S} +  \mathcal{F}_W.
   \label{eq:FI_general}
\end{equation}

To have an inhomogeneous state, in addition to the condition in Eq. (\ref{eq:u_e_cond1}), we require that $\mathcal{F}_{S}(V_S=V) > \mathcal{F}_I$ and $\mathcal{F}_{N}(V_N=V) > \mathcal{F}_I$, i.e. the $S-N$ state is energetically favorable with respect to both the fully superconducting and normal states. By considering the spatial distributions of the superconducting and normal regions $V_{S,E}^T = L_z d_S (2 \lambda_S)$ and $V_{N,E}^T = L_z d_N (2 \lambda_N)$ (see Fig. \ref{fig:wire}), the above inequalities reduce to:
\begin{eqnarray}
    L_z \mathcal{A}_S \Delta F - L_z d_S E^2 (\lambda_S \epsilon_S - \lambda_N \epsilon_N) -F_W &>&0 \nonumber \\
    -L_z \mathcal{A}_N \Delta F + L_z d_N E^2 (\lambda_S \epsilon_S - \lambda_N \epsilon_N) -F_W &>&0.
\end{eqnarray}

We rewrite the superconducting and normal metal areas as $\mathcal{A}_S = \chi\mathcal{A}$ and $\mathcal{A}_N = (1-\chi) \mathcal{A}$, and the exposed lengths as $d_S = \eta L_x$ and $d_N = (1-\eta) L_x$.
It is also convenient to rescale the electric field over the critical electric field [see Eq. (\ref{eq:E_critical_SN})] and defined the ratio $\kappa =\frac{F_W}{V \Delta F}$.
From a physical point of view, the parameters $\chi$, $\eta$ and $\kappa$ are related to the energy of the superconductor (through the area $\mathcal{A}_S$), to the energy stored in the superconductor in presence of an electric field (since $d_S$ is the length of the superconductor exposed to the electric field) and to the superconductor-normal metal interface energy (through $F_W$).
We can now write the above equations as:
\begin{equation}
 \frac{1-\chi +\kappa}{1-\eta} < \Big (\frac{E}{E_C} \Big)^2 < \frac{\chi -\kappa}{\eta}.
\end{equation} 
The necessary condition to have an electric field that can satisfy these inequalities is $\chi > \kappa + \eta$ and gives the spatial dimensions of the region over which the inhomogeneous state is energetically favorable.

A more quantitative description of the distributions of the superconducting and metallic regions would need a microscopic theory, because the present theory is not able to predict the value of $F_W = L_z l \gamma$. In particular, it gives a boundary on the total area $\mathcal{A}_S$ but it is not able to predict the length $l$ of the superconductor-normal interface. In addition, the surface energy $\gamma$ can be estimated only with the aid of a microscopic theory. In fact, the GL formalism (generally used in similar situations) gives  information only when the order parameter can change over the coherence length  $\xi$ \cite{deGennes, tinkham2012introduction}. In the  present experiment $L_y$ and $L_z$ are shorter or of the same order of $\xi$ so that the GL theory does not hold and it fails to predict the superconductor-normal metal surface energy.

Despite these limitations, the simple classical thermodynamical model presented in this work predicts (for certain parameters and electric field) the emergence of an inhomogeneous state induced by the static electric field. The latter is compatible the experimental observations emerging both in the $I-V$ characteristics and in the zero-bias resistance for high values of the gate voltage (i.e. the electric field).

\section{Possible applications} \label{Possible applications}

\begin{figure}
   \begin{center}
    \includegraphics{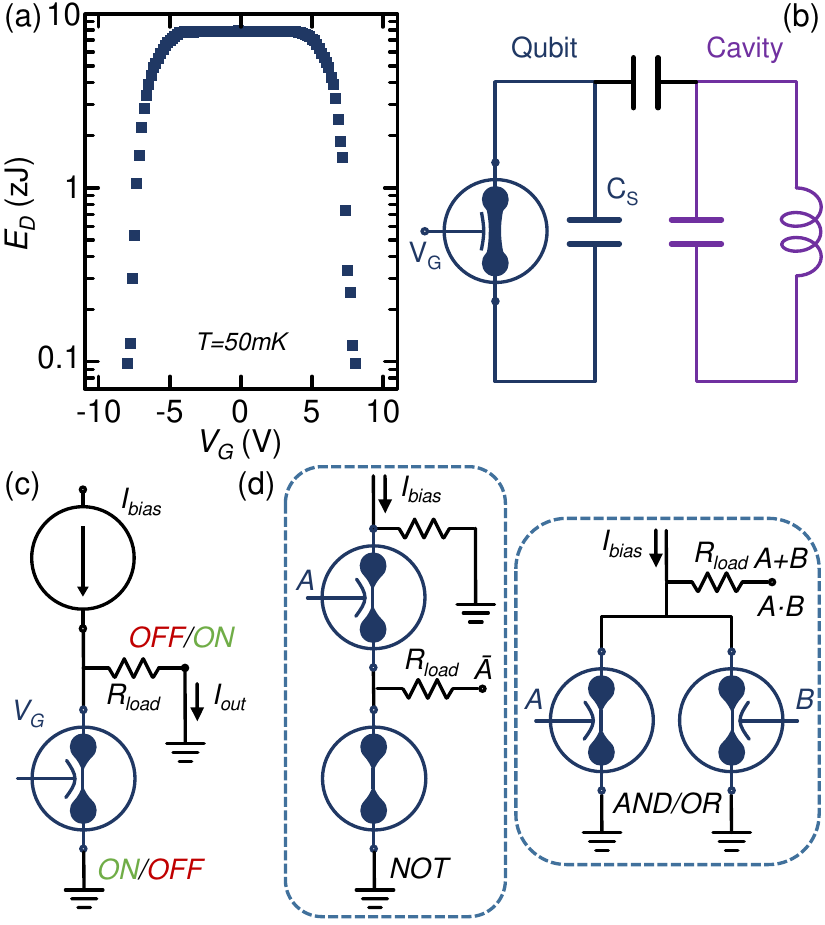}
    \end{center}
    \caption{(a) Josephson coupling energy $E_D$ as a function of the gate voltage $V_G$ for device $2$ at $T=50$mK. (b) Schematic representation of the $Dayemon$ device, where the qubit and the capacitively-coupled resonant cavity are represented. (c) Schematic of the electric-field-controlled cryotron $EF-Tron$. (c) Schematic of classic logic gates based on the $EF-Tron$: $NOT$/$AND$/$OR$. 
      }  
    \label{fig:Fig6}
\end{figure} 

The phenomenology presented in the above experimental sections of this work could pave the way to the design of a number of superconducting devices with gate-tunable behavior and performances, such as  interferometers \cite{Clarke2004, Giazotto2010, Strambini2016}, photon-detectors \cite{Gol2001}, parametric amplifers \cite{Bergeal2010}, coherent caloritronic systems \cite{Giazotto2012, Mart2014, Fornieri2015, Fornieri2017, Perez2014}, metallic gate-tunable transmons ("gatemons") \cite{Larsen2015, deLange2015} and electric-field-controlled cryotrons \cite{McCaughan2014, Zhao2017}. In the following we will focus on the last two aforementioned applications.

The Josephson coupling energy $E_D$ is related to the junction critical current $I_C$ by $E_D=\hbar I_C/2e$, where $\hbar$ is the reduced Planck constant and $e$ is the electron charge. Therefore, the tunability of the critical Josephson current of our Dayem bridge field-effect transistors (see Fig. \ref{fig:Fig1}-b) is reflected in the possibility of controlling $E_D$ across at least two orders of magnitude, as shown in Fig. \ref{fig:Fig6}-a. This capability is nowadays exploited to implement superconducting qubits in the form of gate-tunable semiconductor-based transmons \cite{Larsen2015, deLange2015}, the so-called "gatemons". Similarly our $DB-FET$s coul be the key element for the realization of fully metallic gate-tunable Dayem bridge transmons, the "Dayemons". In such a configuration, the qubit is composed of the $DB-FET$ shunted by a capacitance $C_S$ and is capacitevely coupled to a transmission line resonant cavity \cite{Larsen2015} as depicted in in Fig. \ref{fig:Fig6}-b. If the Josephson energy is far greater than the charging energy ($E_D\gg E_C$), the qubit transition energy takes the form $f_Q\simeq \sqrt{8E_CE_D}/h$, where the charging energy can be calculated from the total capacitance $C_{tot}$ through $E_C=e^2/2C_{tot}$ \cite{Larsen2015}. By assuming $C_{tot}=100$fF we can calculate the expected qubit frequency for different values of the gate voltage applied to our $DB-FET$s. For both the presented the devices the qubit frequency can be tuned of more one order of magnitude by changing $V_G$. In particular, for device $2$ the value of $f_Q$ stays constant for values of gate voltage reaching $4$V [$f_Q(V_G=0-4\mathrm{V})\simeq135$GHz] and then monotonically decreases by further rising $V_G$ [$f_Q(V_G=6.8\mathrm{V})\simeq88$GHz and $f_Q(V_G=7.5\mathrm{V})\simeq28$GHz]. On the one hand,  the "Dayemons" are predicted to show similar performances of conventional semiconductor-based transmons ("gatemons")  \cite{Larsen2015, deLange2015}. On the other hand, the "Dayemons" are a \textit{monolitic} architecture: they are made of a single superconducting metal without any interface between different materials and any oxidation process. Therefore, this is a robust and scalable technology for the implementation of superconducting qubits.

The modulation of the critical current through an external gate electrode is one of the key elements of superconducting classic computation \cite{Likharev2012}. Recently, the challenge has been addressed by the development of nanocryotrons, the so-called $nTron$s, that use an input gate current to tune the supercurrent transport of a metallic channel \cite{McCaughan2014, Zhao2017}. Analogously, the $DB-FET$ can be employed to realize an electric-field-controlled cryotron $EF-Tron$, as depicted in Fig. \ref{fig:Fig6}-c, with the advantage of a very large input-to-output impedance. The bias current is chosen to be lower than the film intrinsic critical current ($I_{bias}<I_{C0}$), therefore for $V_G=0$ all the current flows into the transistor and the output current is zero ($I_{out}=0$). When $I_{bias}>I_C(V_G)$ the channel switches to the normal state and $I_{out}\neq 0$. In our device we can employ $I_{bias}=10\mu\mathrm{A}<I_{C0}$, consequently a load resistance $R_{load}=1\mathrm{k}\Omega \gg R_N$ would provide an output voltage of 10mV. On the one hand, in order to have a sizeable output current the load resistance is required to be much smaller than the normal-state resistance of the transistor ($R_{load}\ll R_N$). Yet, values of $R_{load}$ on the order of the kiloohm are necessary to obtain significant output voltages. As a consequence, transitors with high normal state resistance are requested. This can be achieved by employing structures with channels consituted of long metallic wires (a few $\mu$m) \cite{DeSimoni2018}.

Fig. \ref{fig:Fig6}-d shows the schematic of the realization of the $NOT$/$AND$/$OR$ logic gates with the $EF-Tron$ technology. The $NOT$ gate is formed by the series of two constrictions: the first can be tuned by the input gate electrode ($A$) while the second is characterized by a critical current lower than $I_{bias}$. When the input $A$ is $low$ the current can pass through the first constriction without switching it, while the second constriction passes to the normal state. As a consequence, the current flows through $R_{load}$ and the output signal is $high$. On the contrary, when $A$ is $high$ the first constriction switches to the normal state and the current does not flow in the device core. Therefore, the output signal is $low$. The $AND$ and $OR$ logic gates (see Fig. \ref{fig:Fig6}-d) are realized by placing two $EF-Tron$s in parallel and the discrimination between the two configurations resides in the magnitude of $I_{bias}$ compared to the critical current of the single constriction: if $I_{bias}>I_{CA}=I_{CB}$ the device behaves as an $OR$ logic gate, while if $I_{bias}<I_{CA}=I_{CB}$ the architecture works as an $AND$. 

In this architecture the minimum switching energy for a single logic operation is given by $E_{sw}\simeq LI_{bias}^2$, where $L$ is the device inductance \cite{McCaughan2014}. In our devices $L\simeq1$pH gives $E_{sw}\simeq10^{-22}$ J for the switching of a single bit, which is some orders of magnitude lower than in single flux quantum systems \cite{Volkmann2013}. Furthermore, the switching time estimated from $\tau=L/R_N$ is on the order of $10^{-15}$ seconds that corresponds to switching frequencies of 1000THz. Finally, the $EF-Tron$ technology shows several interesting features for real circuits: strong gate/potential isolation, i.e. the measured gate impedance has been measured to vary from 1 to 1000 TOhm dependening on the used dielectric material \cite{DeSimoni2018, Paolucci2018}, low gate dissipation, and simple \textit{monolitic} fabrication process, structure and composition.

\section{Conclusions}
In conclusion, we showed a broad range set of transport experiments performed on different fully metallic Dayem bridge field-effect transistors and a thermodynamic model which is able to account to some phenomenologies of our findings. 

The measurements of the critical current as a function of gate voltage highlighted the suppression of $I_C$ in a symmetric fashion with the polarity of $V_G$ and the growth of a sub-gap dissipative component just above the full annihilation of supercurrent. Furthermore, the normal-state resistance of our $DB-FET$ turned out to be completely insensitive to electric field. All these results suggest that field-effect induced charge accumulation/depletion is too small to be detected in our experimental conditions and cannot be at the origin of the presented phenomenology. By independently controlling the voltage ($V_{G1}$ and $V_{G2}$) applied to two gate electrodes placed at the opposite sides of the constriction we demonstrated the independence of the effect of $V_{G1}$ and $V_{G2}$ on $I_C$. As a consequence, all this seems to suggest that the above described phenomena occur at the metal surface and affect the superconductor over a few times the coherence length $\xi$. 

The measurements of the Josephson current as a function of temperature showed the transition from a ballistic constriction Kulik-Omelyanchuck behavior to a tunnel-like Ambegaokar-Baratoff characteristic by increasing the gate voltage. This change of behavior is confirmed by fitting the experimental data with a model describing the critical current of a generic Josephson junction. In particular, the effective transmission probability of the constriction extrapolated from the fits drops to zero for values of gate voltage reaching its critical value $V_{GC}$ for the different samples.

The zero-bias resistance measurements highlighted the independence of the superconducting critical temperature and critical magnetic field (onset) on $V_G$ and the growth of a sub-gap resistive component for gate voltages approaching $V_{GC}$ (which is reflected in a widening of the superconducting transition with $B$). These results are in agreement with the $I_C$ vs $V_G$ measurements and suggest that the electric field induces the creation of an inhomogeneous superconductor-normal metal state. The $I_C$ vs $B$ experiments showed two main features: the plateau of the critical current with $V_G$ widens by increasing the out-of-plane magnetic field, while the value of $V_{GC}$ lowers for increasing $B$.

We presented a phenomenological thermodynamic model which is able to account for some of the experimental observations: the electric field-induced suppression of the supercurrent and the emergence of an inhomogeneous normal/superconducting state. The complete understanding of the impact of the electric field on superconductivity would benefit from a set of complementary experiments, such as probing the superconductor density of states through tunnelling spectroscopy, investigating the phase rigidity in superconducting quantum interference devices, examining the thermal transport in phase-biased Josephson junctions, and studying the kinetic inductance in superconducting resonators. 

From an application point of view, this phenomenology paves the way to the realization of gate-tunable superconducting devices such as interferometers  \cite{Clarke2004, Giazotto2010, Strambini2016}, photon-detectors \cite{Gol2001}, coherent caloritronic systems  \cite{Giazotto2012, Mart2014, Fornieri2015, Fornieri2017, Perez2014}, parametric amplifers \cite{Bergeal2010}, gate-tunable metallic superconducting qubits \cite{Larsen2015, deLange2015} and superconducting classic electronics \cite{McCaughan2014, Zhao2017}. Finally, the use of higher critical temperature or resistivity metallic superconductors, such as vanadium and niobium, would boost the implementation of our technology in high-speed and low-dissipation superconducting electronics.

\begin{acknowledgments}
The authors acknowledge the European Research Council under the European Unions Seventh Framework Programme (FP7/2007-2013)/ERC Grant No. 615187 - COMANCHE, the European Union (FP7/2007-2013)/REA Grant No. 630925 - COHEAT and the MIUR under the FIRB2013 Project Coca (Grant No. RBFR1379UX) for partial financial support. The work of G.D.S. and F.P. was partially funded by Tuscany Region under the FARFAS 2014 project SCIADRO. The work of F.P. work has been partially supported by the Tuscany Government, POR FSE 2014 -2020, through the INFN-RT2 172800 Project. A.B. acknowledges the CNR-CONICET cooperation program “Energy conversion in quantum nanoscale hybrid devices” and the Royal Society through the International Exchanges between the UK and Italy (Grant No. IES R3 170054).
\end{acknowledgments}

\end{document}